# Review of the Safety of LHC Collisions

**LHC Safety Assessment Group**

**John Ellis[1], Gian Giudice[1], Michelangelo Mangano[1],
Igor Tkachev[2] and Urs Wiedemann[1]**

[1]Theory Division, Physics Department, CERN,
CH 1211 Geneva 23, Switzerland
[2]Institute for Nuclear Research of Russian Academy of Sciences,
Moscow 117312, Russia

**Abstract**

*The safety of collisions at the Large Hadron Collider (LHC) was studied in 2003 by the LHC Safety Study Group, who concluded that they presented no danger. Here we review their 2003 analysis in light of additional experimental results and theoretical understanding, which enable us to confirm, update and extend the conclusions of the LHC Safety Study Group. The LHC reproduces in the laboratory, under controlled conditions, collisions at centre-of-mass energies less than those reached in the atmosphere by some of the cosmic rays that have been bombarding the Earth for billions of years. We recall the rates for the collisions of cosmic rays with the Earth, Sun, neutron stars, white dwarfs and other astronomical bodies at energies higher than the LHC. The stability of astronomical bodies indicates that such collisions cannot be dangerous. Specifically, we study the possible production at the LHC of hypothetical objects such as vacuum bubbles, magnetic monopoles, microscopic black holes and strangelets, and find no associated risks. Any microscopic black holes produced at the LHC are expected to decay by Hawking radiation before they reach the detector walls. If some microscopic black holes were stable, those produced by cosmic rays would be stopped inside the Earth or other astronomical bodies. The stability of astronomical bodies constrains strongly the possible rate of accretion by any such microscopic black holes, so that they present no conceivable danger. In the case of strangelets, the good agreement of measurements of particle production at RHIC with simple thermodynamic models constrains severely the production of strangelets in heavy-ion collisions at the LHC, which also present no danger.*

# 1 - Introduction

The controlled experimental conditions offered by accelerators permit the detailed study of many natural phenomena occurring in the Universe. Many of the fundamental particles, such as muons, pions and strange particles, were first discovered among the cosmic rays, and were subsequently studied with accelerators. As the energies of accelerators have increased, they have also revealed many heavier and less stable particles, such as those containing heavier quarks, as well as the carrier particles of the weak interactions. Though not present in ordinary stable matter, these particles played important roles in the early history of the Universe, and may still be important today in energetic astronomical bodies such as those producing the cosmic rays.

The energies of accelerators have been increasing regularly over the past decades, though they are still far below those of the most energetic cosmic rays. With each increase in accelerator energy, one may ask whether there could be any risk associated with new phenomena that might be revealed. It has always been reassuring that higher-energy cosmic rays have been bombarding the Earth since its creation with no disastrous side-effects. On the other hand, the particles produced in cosmic-ray collisions typically have different velocities with respect to the Earth from those produced by accelerators, so the circumstances are not directly comparable. The question has therefore been asked whether Earth's immunity to cosmic-ray collisions also applies to accelerator collisions.

The Large Hadron Collider (LHC) accelerator is nearing completion at CERN. It is designed to collide pairs of protons each with energies of 7 TeV (somewhat more than 7000 times the rest mass-energy of the proton), and pairs of lead nuclei each with energies of about 2.8 TeV per proton or neutron (nucleon). Though considerably higher than the energies of previous accelerators, these energies are still far below those of the highest-energy cosmic-ray collisions that are observed regularly on Earth. In light of safety questions about previous accelerators, and in advance of similar questions about the LHC, the CERN management requested a report on the safety of the LHC by the LHC Safety Study Group, a panel of independent experts, which was published in 2003 [1]. This report concluded that there is no basis for any conceivable threat from the LHC.

The advent of LHC operations now revives interest in safety questions, so the CERN management has commissioned us to review the arguments presented in the 2003 report and previous studies of the possible production of new particles, and to update them in light of experimental results from the Brookhaven relativistic heavy-ion collider (RHIC), in particular, as well as of recent theoretical speculations about new phenomena.

We consider all the speculative scenarios for new particles and states of matter that have been discussed in the scientific literature and raise potential safety issues. Our methodology is based on empirical reasoning using experimental observations, and hence could be extended to other exotic phenomena that might be cause for concerns in the future.

We focus our attention mainly on two phenomena of current interest, namely the possible production of microscopic black holes, such as might appear in certain theoretical models featuring additional dimensions of space, and the possible production of 'strangelets', hypothetical pieces of matter analogous to conventional nuclei, but containing also many of the heavier strange quarks. These were both considered carefully in the 2003 report of the LHC Safety Study Group [1]. Their conclusions for strangelets, obtained in connection with heavy-ion collisions, apply with equal or greater force to proton-proton collisions at the LHC.

In the case of microscopic black holes, there has been much theoretical speculation since 2003 about their existence and their possible experimental signatures, as reviewed in [2], where references may be found. In the case of strangelets, detailed experimental measurements at the Brookhaven National Laboratory's Relativistic Heavy-Ion Collider (RHIC) of the production of particles containing different numbers of strange quarks [3] enable one to refine previous arguments that, in the event they exist, strangelets would be less likely to be produced at the LHC than at RHIC.

Before discussing these two hypothetical phenomena in more detail, we first review in Section 2 estimates of the rates of collisions of high-energy cosmic rays with different astronomical bodies, such as the Earth, Sun, and others. We estimate that the Universe is replicating the total number of collisions to be made by the LHC over $10^{13}$ times per second, and has already done so some $10^{31}$ times since the origin of the Universe. The fact that astronomical bodies withstand cosmic-ray bombardment imposes strong upper limits on many hypothetical sources of danger. In particular, as we discuss in Section 3, neither the creation of vacuum bubbles nor the production of magnetic monopoles at the LHC is a case for concern.

In the case of the hypothetical microscopic black holes, as we discuss in Section 4, if they can be produced in the collisions of elementary particles, they must also be able to decay back into them. Theoretically, it is expected that microscopic black holes would indeed decay via Hawking radiation, which is based on basic physical principles on which there is general consensus. If, nevertheless, some hypothetical microscopic black holes should be stable, we review arguments showing that they would be unable to accrete matter in a manner dangerous for the Earth [2]. If some microscopic black holes were produced by the LHC, they would also have been produced by cosmic rays and have stopped in the Earth or some other astronomical body, and the stability of these astronomical bodies means that they cannot be dangerous.

In the case of the equally hypothetical strangelets, we review in Section 5 the data accumulated at RHIC on the abundances and velocities of strongly-interacting particles, including those containing one or more strange quarks, produced at RHIC and in previous heavy-ion collision experiments [3]. All these data are very consistent with a simple thermodynamic production mechanism that depends only on the effective temperature and the net density of baryons (nucleons). The effective temperature agrees well with first-principles theoretical calculations, and the net density of baryons decreases as the energy increases, again in agreement with theoretical calculations. Calculations for heavy-ion collisions at the LHC give a similar effective temperature and a lower net density of baryons than at RHIC. This means that the LHC could only produce strangelets at a lower rate, if they exist at all.

We conclude by reiterating the conclusion of the LHC Safety Group in 2003 [1]: there is no basis for any conceivable threat from the LHC. Indeed, theoretical and experimental developments since 2003 have reinforced this conclusion.

## 2 – The LHC compared with Cosmic-Ray Collisions

The LHC is designed to collide two counter-rotating beams of protons or heavy ions. Proton-proton collisions are foreseen at an energy of 7 TeV per beam. An equivalent energy in the centre of mass would be obtained in the collision of a cosmic-ray proton with a fixed target such as the Earth or some other astronomical body if its energy reaches or exceeds $10^8$ GeV, i.e., $10^{17}$ eV [4]. When the LHC attains its design collision rate, it will produce about a billion proton-proton collisions per second in each of the major detectors ATLAS and CMS. The effective amount of time each year that the LHC will produce collisions at this average luminosity is about ten million seconds. Hence, each of the two major detectors is expecting to obtain about $10^{17}$ proton-proton collisions over the planned duration of the experiments.

As seen in Fig. 1, the highest-energy cosmic rays observed attain energies of around $10^{20}$ eV, and the total flux of cosmic rays with energies of $10^{17}$ eV or more that hit each square centimeter of the Earth's surface is measured to be about $5 \times 10^{-14}$ per second [5]. The area of the Earth's surface is about $5 \times 10^{18}$ square centimeters, and the age of the Earth is about 4.5 billion years. Therefore, over $3 \times 10^{22}$ cosmic rays with energies of $10^{17}$ eV or more, equal to or greater than the LHC energy, have struck the Earth's surface since its formation. This means [6] that Nature has already conducted the equivalent of about a hundred thousand LHC experimental programmes on Earth already – and the planet still exists.

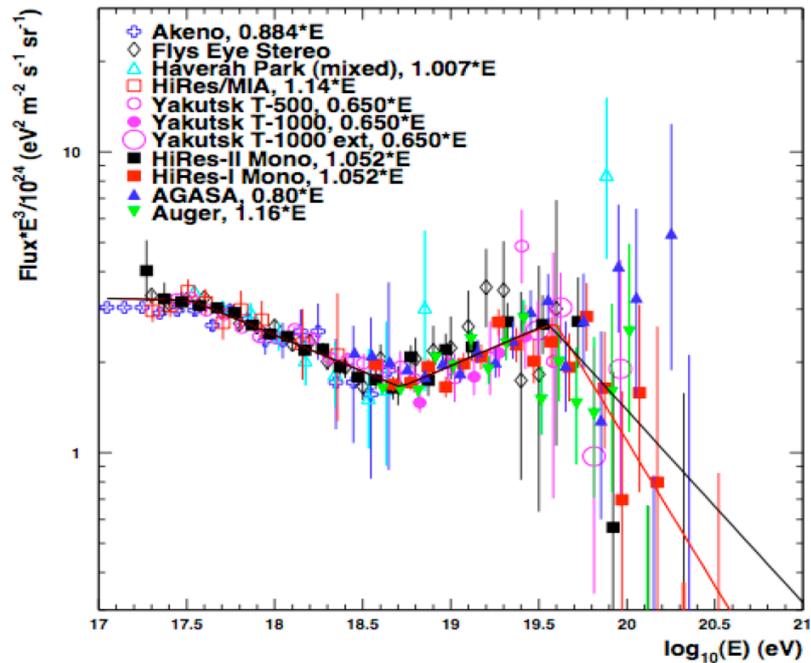

*Fig. 1: The spectrum of ultra-high-energy cosmic rays, as measured by several experiments [5]. Every cosmic ray with an energy shown in this plot, namely above $10^{17}$ eV, liberates in its collision with the atmosphere more energy in its centre-of-mass frame than does a proton-proton collision at the LHC.*

Other astronomical bodies are even larger. For example, the radius of Jupiter is about ten times that of the Earth, and the radius of the Sun is a factor of ten larger still. The surface area of the Sun is therefore 10,000 times that of the Earth, and Nature has therefore already conducted the LHC experimental programme about one billion times [6] via the collisions of cosmic rays with the Sun – and the Sun still exists.

Moreover, our Milky Way galaxy contains about $10^{11}$ stars with sizes similar to our Sun, and there are about $10^{11}$ similar galaxies in the visible Universe. Cosmic rays have been hitting all these stars at rates similar to collisions with our own Sun. This means that Nature has already completed about $10^{31}$ LHC experimental programmes since the beginning of the Universe. Moreover, each second, the Universe is continuing to repeat about $3 \times 10^{13}$ complete LHC experiments. There is no indication that any of these previous "LHC experiments" has ever had any large-scale consequences. The stars in our galaxy and others still exist, and conventional astrophysics can explain all the astrophysical black holes detected.

Thus, the continued existence of the Earth and other astronomical bodies can be used to constrain or exclude speculations about possible new particles that might be produced by the LHC.

### 3 – Vacuum Bubbles and Magnetic Monopoles

These large rates for the collisions of cosmic rays at energies higher than the LHC imply directly that there can be no danger to the Earth from the production of bubbles of new vacuum or magnetic monopoles at the LHC [1].

It has often been suggested that the Universe might not be absolutely stable, because the state that we call the 'vacuum' might not be the lowest-energy state. In this case, our 'vacuum' would eventually decay into such a lower-energy state. Since this has not happened, the lifetime before any such decay must be longer than the age of the Universe. The possible concern about high-energy particle collisions is that they might stimulate the production of small 'bubbles' of such a lower-energy state, which would then expand and destroy not just the Earth, but potentially the entire Universe.

However, if LHC collisions could produce vacuum bubbles, so also could cosmic-ray collisions. This possibility was first studied in [7], and the conclusions drawn there were reiterated in [8]. These bubbles of new vacuum would have expanded to consume large parts of the visible Universe several billion years ago already. The continued existence of the Universe means that such vacuum bubbles are not produced in cosmic-ray collisions, and hence the LHC will also not produce any vacuum bubbles.

There have also been suggestions over the years that there might exist magnetic monopoles, particles with non-zero free magnetic charge. As was originally pointed out by Dirac, any free magnetic charge would be quantized, and necessarily much larger in magnitude than the electric charge of the electron or proton. For this reason, searches for magnetic monopoles have looked for heavily-ionizing particles as well as for quanta of magnetic charge, and this search will be continued at the LHC.

In some grand unified theories, though not in the Standard Model of particle physics, magnetic monopoles might also catalyze nucleon decay, by transforming protons and neutrons into electrons or positrons and unstable mesons. In this case, successive collisions with large numbers of nuclei would release considerable energy. The magnetic monopoles that might have such properties are expected to weigh $10^{15}$ GeV or more, far too heavy to be produced at the LHC. Nevertheless, here we consider the possibility of producing light proton-eating magnetic monopoles at the LHC.

A quantitative discussion of the impact of such magnetic monopoles on Earth was presented in [1], where it was concluded that only a microgram of matter would be destroyed before the monopole exited the Earth. Independently of this conclusion, if monopoles could be produced by the LHC, high-energy cosmic rays would already have created many of them when striking the Earth and other astronomical bodies. Since they would have large magnetic charges, any monopoles produced by cosmic rays would have been stopped by the material of the Earth [2]. The continued existence of the Earth and other astronomical bodies after billions of years of high-energy cosmic-ray bombardment means that any monopoles produced could not catalyze proton decay at any appreciable rate. If the collisions made in the LHC could produce dangerous monopoles, high-energy cosmic rays would already have done so.

The continued existences of the Earth and other astronomical bodies such as the Sun mean that any magnetic monopoles produced by high-energy cosmic rays must be harmless. Likewise, if any monopoles are produced at the LHC, they will be harmless.

**4 – Microscopic Black Holes**

As has already been discussed, the LHC will make collisions with a much lower centre-of-mass energy than some of the cosmic rays that have been bombarding the Earth and other astronomical bodies for billions of years. We estimate that, over the history of the Universe, Nature has carried out the equivalent of $10^{31}$ LHC projects (defined by the integrated luminosity for cosmic-ray collisions at a centre-of-mass energy of 14 TeV or more), and continues to do so at the rate of over $10^{13}$ per second, via the collisions of energetic cosmic rays with different astronomical bodies.

There is, however, one significant difference between cosmic-ray collisions with a body at rest and collisions at the LHC, namely that any massive new particles produced by the LHC collisions will tend to have low velocities, whereas cosmic-ray collisions would produce them with high velocities. This point has been considered in detail [2] since the 2003 report by the LHC Safety Study Group [1]. As we now discuss, the original conclusion that LHC collisions present no dangers is validated and strengthened by this more recent work.

We recall that the black holes observed in the Universe have very large masses, considerably greater than that of our Sun. On the other hand, each collision of a pair of protons in the LHC will release an amount of energy comparable to that of two colliding mosquitos, so any black hole produced would be much smaller than those known to astrophysicists. In fact, according to the conventional gravitational theory of General Relativity proposed by Einstein, many of whose predictions have subsequently been verified, there is no chance that any black holes could be produced at the LHC, since the conventional gravitational forces between fundamental particles are too weak.

However, there are some theoretical speculations that, when viewed at very small distances, space may reveal extra dimensions. In some such theories, it is possible that the gravitational force between pairs of particles might become strong at the energy of the LHC.

As was pointed out 30 years ago by Stephen Hawking [9], it is expected that all black holes are ultimately unstable. This is because of very basic features of quantum theory in curved spaces, such as those surrounding any black hole. The basic reason is very simple: it is a consequence of quantum mechanics that particle-antiparticle pairs must be created near the event horizon surrounding any black hole. Some particles (or antiparticles) disappear into the black hole itself, and the corresponding antiparticles (or particles) must escape as radiation. There is broad consensus among physicists on the reality of Hawking radiation, but so far no experiment has had the sensitivity required to find direct evidence for it.

Independently of the reasoning based on Hawking radiation, if microscopic black holes were to be singly produced by colliding the quarks and gluons inside protons, they would also be able to decay into the same types of particles that produced them [10]. The reason being that in this case they could not carry any conserved quantum number that is not already carried by the original quarks and gluons, and their decay back to the initial state partons would be allowed. For this reason, a microscopic black hole cannot be completely black. In standard quantum physics, the decay rate would be directly related to the production rate, and the expected lifetime would be very short. The case of pair production of black holes carrying new and opposite conserved quantum numbers leads to similar conclusions: only their ground state is guaranteed to be stable, and any further accretion of normal matter in the form of quarks, gluons or leptons would immediately be radiated away. Both this and the existence of Hawking radiation are valid in the extra-dimensional scenarios used to suggest the possible production of microscopic black holes.

One might nevertheless wonder what would happen if a stable microscopic black hole could be produced at the LHC [2]. However, we reiterate that this would require a violation of some of the basic principles of quantum mechanics – which is a cornerstone of the laws of Nature – in order for the black hole decay rate to be suppressed relative to its production rate, and/or of general relativity – in order to suppress Hawking radiation.

Most black holes produced at the LHC or in cosmic-ray collisions would have an electric charge, since they would originate from the collisions of charged quarks. A charged object interacts with matter in an experimentally well-understood way. A direct consequence of this is that charged and stable black holes produced by the interactions of cosmic rays with the Earth or the Sun would be slowed down and ultimately stopped by their electromagnetic interactions inside these bodies, in spite of their initial high velocities. The complete lack of any macroscopic effect caused by stable black holes, which would have accumulated in the billions during the lifetime of the Earth and the Sun if the LHC could produce them, means that either they are not produced, or they are all neutral and hence none are stopped in the Earth or the Sun, or have no large-scale effects even if they are stopped.

If a black hole were to be produced by a cosmic ray, as it traveled through the Earth it would absorb preferentially similar numbers of protons and neutrons, because their masses are larger than that of the electron. It would, therefore, develop and maintain a positive charge, even if it were produced with no electric charge. The standard neutralization process due to the quantum creation of particle-antiparticle pairs near the horizon – the Schwinger mechanism – relies on principles very similar to those at the basis of Hawking radiation, and would likely not operate if the latter was suppressed. Thus, combining the hypotheses that black holes are simultaneously neutral and stable and accrete matter requires some further deviation from basic physical laws. There is no concrete example of a consistent scenario for microphysics that would exhibit such behaviour. Furthermore, it is possible [2] to exclude any macroscopic consequences of black holes even if such unknown mechanisms were realized, as we now discuss.

The rate at which any stopped black hole would absorb the surrounding material and grow in mass is model-dependent. This is discussed in full detail in [2], where several accretion scenarios, based on well-founded macroscpic physics, have been used to set conservative, worst-case-scenario limits to the black hole growth rates in the Earth and in denser bodies like white dwarfs and neutron stars. In the extra-dimensional scenarios that motivate the existence of microscopic black holes (but not their stability), the rate at which absorption would take place would be so slow if there are seven or more dimensions that Earth would survive for billions of years before any harm befell it. The reason is that in such scenarios the size of the extra dimensions is very small, so small that the evolution driven by the strong extra-dimensional gravity forces terminates while the growing black hole is still of microscopic size. If there are only five or six dimensions of space-time relevant at the LHC scale, on the other hand, the gravitational interactions of black holes are strong enough that their impact, should they exist, would be detectable in the Universe.

In fact, ultra-high-energy cosmic rays hitting dense stars such as white dwarfs and neutron stars would have produced black holes copiously during their lifetimes. Such black holes, even if neutral, would have been stopped by the material inside such dense stars. The rapid accretion due to the large density of these bodies, and to the strong gravitational interactions of these black holes, would have led to the destruction of white dwarfs and neutron stars on time scales that are much shorter than their observed lifetimes [2]. The final stages of their destruction would have released explosively large amounts of energy, that would have been highly visible. The observation of white dwarfs and neutron stars that would have been destroyed in this way tells us that cosmic rays do not produce such black holes, and hence neither will the LHC.

To conclude: in addition to the very general reasoning excluding the possibility that stable black holes exist, and in particular that they could only be neutral, we therefore have very robust empirical evidence either disproving their existence, or excluding any consequence of it.

## 5 - Strangelets

The research programme of the CERN Large Hadron Collider also includes the collisions of ultra-relativistic lead and other nuclei (ions). The main scientific goal of this heavy-ion programme is to produce matter at the highest temperatures and densities attainable in the laboratory, and to study its properties. This programme is expected to produce, in very small quantities, primordial plasma of the type that filled the Universe when it was about a microsecond old.

The normal matter of which we are made, and which constitutes all the known visible matter in the Universe, is composed of the two lightest types of quarks, the up and down quarks. Heavier, unstable quarks have been discovered in cosmic-ray collisions and at accelerators, and the lightest of these is the strange quark. Particles containing strange quarks have been produced regularly in the laboratory for many decades, and are known to decay on time scales of the order of a nanosecond, or faster. Such lifetimes are characteristic of the weak interaction responsible for radioactivity, which governs their decay. Some unstable particles containing two or three strange quarks have also been observed. Particles including one strange quark have been shown to bind to nuclei, the so-called hypernuclei, which are however unstable and promptly decay, again with nanosecond time scales. Apart from rapidly decaying nuclei with two particles each containing one strange quark [11], no nuclei containing multiple strange quarks are known.

Strange quark matter is a hypothetical state of matter, which would consist of large, roughly equal numbers of up, down and strange quarks. Hypothetical small lumps of strange quark matter, having atomic masses comparable to ordinary nuclei, are often referred to as strangelets. As discussed in more detail in the Appendix, most theoretical studies of strangelets conclude that, if they exist, they must be unstable, decaying with a typical strange-particle lifetime of around a nanosecond. In this case, any production of strangelets would pose no risk. However, it has been speculated that strange quark matter might weigh less than conventional nuclear matter with the same number of up and down quarks, but not for atomic numbers smaller than 10. In this very hypothetical case, such a strangelet would be stable. It has been further speculated that, if produced, strangelets could coalesce with normal matter and catalyze its conversion into strange matter, thereby creating an ever-growing strangelet. This hypothetical scenario underlies concerns about strangelet production at accelerators, which were discussed previously in [8] and [1].

It is generally expected that any stable strangelet would have a positive charge, in which case it would be repelled by ordinary nuclear matter, and hence unable to convert it into strange matter [8], see [12], however. In some model studies, one finds that negatively-charged strangelets can also exist, but are unstable since the positively-charged states have lower energy [13]. However, there is no rigorous proof that the charge of a stable strangelet must be positive, nor that a negatively-charged strangelet cannot be metastable, i.e., very long-lived. So, one should also consider the possibility of a negatively-charged stable or very long-lived strangelet.

Prior to the start of the Relativistic Heavy-Ion Collider (RHIC), a study was carried out [8] to assess hypothetical scenarios for the production of strangelets in heavy-ion collisions. Additional arguments were given in [14], and a reassessment of such a possibility was given in the 2003 Report of the LHC CERN Safety Study Group [1]. We revisit here this topic in light of recent advances in our understanding of the theory and experiment of heavy-ion collisions. These enable us to update and strengthen the previous conclusions about hypothetical scenarios based on strangelet production. More details of our considerations on strangelet production at the LHC are given in the appendix.

The 2003 report summarized the status of direct experimental searches and of theoretical speculations about hypothetical strangelet production mechanisms [1]. More recently, additional direct upper limits on strangelet production have been provided by experimental searches at RHIC [15] and among cosmic rays [16], which have not yielded any evidence for the existence of strangelets. In the near future, additional experimental information may be expected from strangelet searches in samples of lunar soil and from particle detectors in outer space [17].

On the theoretical side, the 2003 report considered three mechanisms for strangelet production [1]: i) a thermal mechanism [3], in which particles are produced as if from a heat bath in thermal equilibrium, ii) a coalescence mechanism, in which particles produced in a heavy-ion collision might combine at late times to form a strangelet, and iii) a distillation mechanism [18], which was proposed as a specific model for strangelet production. According to this last mechanism, a hot quark-gluon plasma with large net baryon number is produced in heavy-ion collisions, and is enriched in strangeness as it cools down by emitting predominantly particles containing strange antiquarks.

As discussed in the Appendix, no evidence has been found in the detailed study of heavy-ion collisions at RHIC for an anomalous coalescence mechanism. In particular, the production rate of light nuclei measured in central Au+Au collisions at RHIC [14], is consistent with the coalescence rates, used in the 2003 Report of the LHC CERN Safety Study Group [1] to rule out strangelet production. There is also considerable experimental evidence against the distillation mechanism. For this mechanism to be operational, the produced matter should have a long lifetime and a large net nucleon density. However, experiments at RHIC confirm the general expectations that the net nucleon density is small and decreases at higher collision energies. Moreover, the plasma produced in the collision is very short-lived, expanding rapidly at about half the velocity of light, and falling apart within $10^{-23}$ seconds [19]. Furthermore, no characteristic difference has been observed in the production of particles containing strange quarks and antiquarks. Hence, a distillation mechanism capable of giving rise to strangelet production is not operational in heavy-ion collisions at RHIC, and this suggestion for strange-particle production has been abandoned for the LHC. On the other hand, as reviewed below, RHIC data strongly support models that describe particle production as emission from a high-temperature heat bath [3].

If they exist, strangelets would be bound states that would be formed initially with an atomic number comparable to that of normal nuclei. Like normal nuclei, strangelets would also contain a significant baryon number. We know from the basic principles of quantum mechanics that, for a strangelet to be formed, its constituents must be assembled in a configuration that contains less than its characteristic binding energy. If this were not the case, the forces between the constituents would not be strong enough to hold them together, and the strangelet would not form. As a consequence, strangelet formation is less likely if the constituents have initially more kinetic energy, and specifically if they emerge from a hotter system. Correspondingly, strangelet production is less likely in a hotter system.

The energy needed to break up a strangelet is similar to that needed to break up a normal nucleus, which is of the order of one to a few million electron volts. Similar energies would be reached in a heat bath with a temperature of ten to several tens of billions of degrees Celsius. However, heavy-ion collisions are known to produce heat baths that are far hotter, reaching temperatures exceeding 1 trillion degrees Celsius [3]. Basic thermodynamics would require most strangelets to melt in such a heat bath, i.e., dissociate into the known strange particles that decay within a nanosecond. For this reason, the likelihood of strangelet production in relativistic heavy-ion collisions can be compared to the likelihood of producing an icecube in a furnace.

The analogy of heavy-ion collisions with a particle furnace has been supported by many detailed measurements in accelerator collisions of the production of different types of particles, including those containing one, two or three strange quarks. Fig. 2 shows one such piece of evidence: the relative rates at which particles are produced in heavy-ion collisions at RHIC is in line with a theoretical calculation assuming a furnace with a temperature around 1.6 trillion degrees [3]. All particle ratios are well-described, including rare particles like the Omega baryon, which contains three strange quarks and which – if compared to the most abundant particles such as pions - is produced only at the per-mille level (see Fig. 2). Supplementary information is provided in the Appendix.

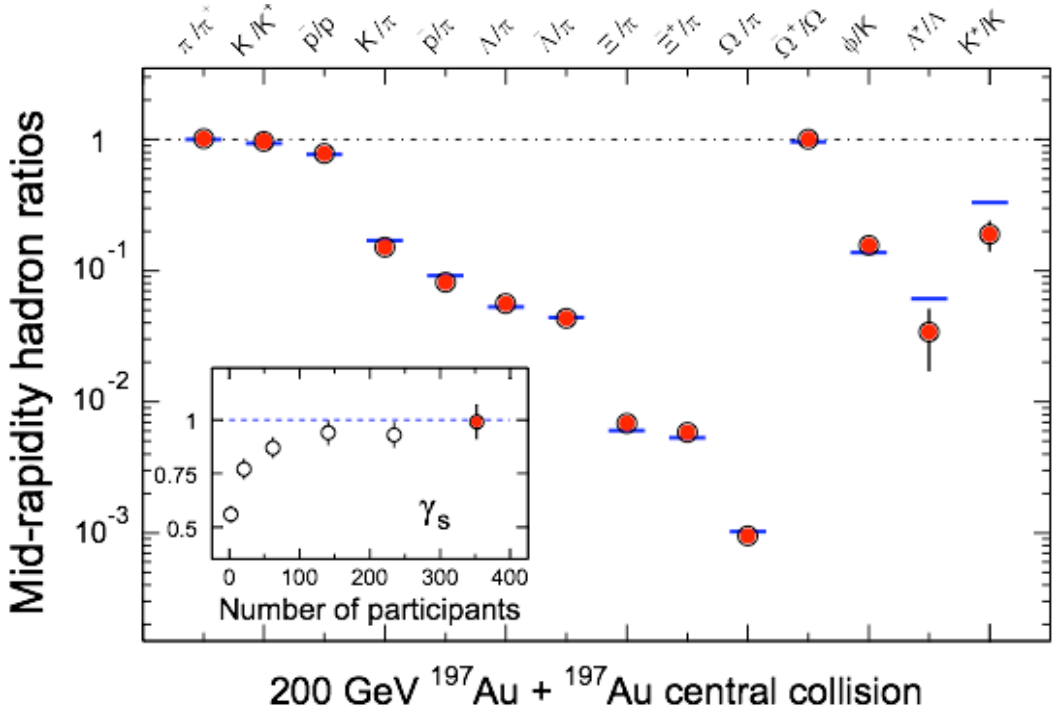

*Fig. 2: The relative amounts of different particles and antiparticles produced at RHIC in the collisions of Gold nuclei with energies $\sqrt{s_{NN}}$ = 200 GeV per nucleon-nucleon collision. All the measurements (red points) agree very well with a simple thermal model (blue lines) with an effective temperature around 1.6 trillion degrees, in line with theoretical calculations, and a net nucleon (quark) density that is lower than in previous, lower-energy experiments. The inset shows that the fraction of strange quarks has saturated at the same density as up and down quarks. Figure taken from refs. [3,23].*

The total number of heavy-ion collisions created at the LHC will be comparable to the total number of heavy ion collisions created at RHIC. The LHC will be at least as hot a furnace as RHIC, in the sense that the systems produced in heavy-ion collisions at the LHC will have an effective temperature that is similar to that produced at RHIC. This is one factor that makes strangelet production no more likely at the LHC than at RHIC. Another major factor pointing in the same direction is that the net density of nucleons, measured by the baryon number, will be lower at the LHC than at RHIC. This is because the system produced in heavy-ion collisions at the LHC is spread over a larger rapidity range, and the same total net baryon number will be spread over a larger volume. As discussed in more detail in the Appendix, this effect has already been seen at RHIC, where the net density of nucleons is lower than in lower-energy experiments, and this trend will continue at the LHC [3]. Since strangelets require baryon number to be formed, this effect makes strangelet production less likely at the LHC than at RHIC.

We conclude on general physical grounds that heavy-ion collisions at the LHC are less likely to produce strangelets than the lower-energy heavy-ion collisions already carried out in recent years at RHIC, just as strangelet production at RHIC was less likely than in previous lower-energy experiments carried out in the 1980s and 1990s [8].

Knowing that strangelet production at the LHC is less likely than at previous lower-energy machines, we now review the arguments that strangelet production in previous lower-energy experiments did not pose any conceivable risk.

It has been shown that the continuing survival of the Moon under cosmic-ray bombardment ensures that heavy-ion collisions do not pose any conceivable threat via strangelet production [8]. This is because cosmic rays have a significant component of heavy ions, as does the surface of the Moon. Since the Moon, unlike planets such as the Earth, is not protected by an atmosphere, cosmic rays hitting the Moon have produced heavy-ion collisions over billions of years at energies that are comparable to or exceed those reached in man-made experiments.

The conclusion reached in [8] required two well-motivated assumptions. Since high-energy cosmic rays include many iron nuclei, which are also prevalent in the Moon's surface, it was assumed that the conditions reached in iron-iron collisions are comparable to those reached in the collisions of gold ions or lead ions that had been studied previously in the laboratory. Secondly, since RHIC and LHC experiments take place in the centre-of-mass reference frame, whereas in cosmic-ray collisions the centre-of-mass frame is moving at high speed, it was necessary to make some assumption about the velocity distribution of any strangelets produced. We recall that high-velocity strangelets might well be broken up by lunar matter before becoming slow enough to coalesce with it.

Since the appearance of [8], the RHIC heavy-ion programme has also studied the collisions of the copper ions, which are closely comparable to iron-iron collisions. The abundances of particles produced in these collisions are described by the same thermal model of a particle furnace that accounts successfully for particle production in gold-gold collisions. Moreover, the velocity distributions of all particle species observed at RHIC are similar to or broader than the distribution assumed in [8]. These observations support the assumptions made in [8], and therefore strengthen their conclusions.

An independent safety argument, which does not require any assumption about the velocity distribution of any hypothetical strangelets, has been given in [20]. The rate of cosmic-ray heavy-ion collisions in interstellar space is known. If these collisions produced any strangelets, these would have accreted in stars and any large-scale coalescence would have resulted in stellar explosions that have not been seen. This complementary argument does, however, assume that any strangelets produced do not decay on a time scale much shorter than that of star formation.

We close this section by summarizing that the successful description of heavy-ion collisions as a particle furnace with a net density of baryons that decreases at higher energies implies that strangelet production at the LHC is less likely than at lower-energy machines (see appendix). The arguments given previously for the safety of lower-energy collisions are strengthened by recent observations at RHIC. Furthermore, we note that the analogy of the LHC with a hot particle furnace will be monitored from the earliest days of heavy-ion collisions at the LHC. A thousand heavy-ion collisions would already suffice for a first test of the thermal model which describes heavy-ion collisions as a particle furnace. This will be among the first data analyses done in the LHC heavy-ion programme, and will immediately provide an experimental confirmation of the basic assumptions on which the safety argument is based.

**6 – Conclusions**

Having reviewed the theoretical and experimental developments since the previous safety report was published, we confirm its findings. There is no basis for any concerns about the consequences of new particles or forms of matter that could possibly be produced by the LHC.

In the case of phenomena, such as vacuum bubble formation via phase transitions or the production of magnetic monopoles, which had already been excluded by the previous report [1], no subsequent development has put into question those firm conclusions. Stable and neutral black holes, in addition to being excluded by all known theoretical frameworks, are either excluded by the stability of astronomical bodies, or would accrete at a rate that is too low to cause any macroscopic effects on timescales much longer than the natural lifetime of the solar system. The previous arguments about the impossibility to produce strangelets at the LHC are confirmed and reinforced by the analysis of the RHIC data.

We have considered all the proposed speculative scenarios for new particles and states of matter that currently raise safety issues. Since our methodology is based on empirical reasoning based on experimental observations, it would be applicable to other exotic phenomena that might raise concerns in the future.

# Appendix

In Section 5 we state that there is no basis for any conceivable threat from strangelet production at the LHC. Our arguments follow closely the line of reasoning of previous reports [8], but they put particular emphasis on two observations. First, on general grounds, the probability for strangelet production decreases with increasing center-of-mass energy. As a consequence, strangelet production at LHC is less likely than at RHIC, just as it was less likely at RHIC than in the heavy-ion programs at lower center-of-mass energies pursued in the 1980s and 1990s. Secondly, RHIC data strongly disfavour models of particle production which were advocated as production mechanisms for strangelets. On the contrary, RHIC data give strong support to a thermal model of particle production, which puts tight upper bounds on strangelet production. In this Appendix, we provide background information to support these statements and the main conclusions drawn from them. In particular, we recall the main arguments of the safety reports [8], and we discuss how these arguments can be strengthened in the light of recent data from RHIC.

**Strangelet properties**

Strange quark matter is a hypothetical state of matter consisting of roughly equal numbers of up, down and strange quarks. It has been speculated that strange quark matter might constitute the true ground state of baryonic matter, being more stable than ordinary nuclei [21,22]. Hypothetical small lumps of strange quark matter, having atomic masses comparable to ordinary nuclei, are often referred to as strangelets. Such strangelets might be either stable or metastable. At present, a first principle theory of strange quark matter is not within theoretical reach. It would require major theoretical breakthroughs in the application of QCD to finite density and to mesoscopic systems. As a consequence, theoretical studies on whether strangelets can exist for some parameter range depend on model-dependent assumptions. As reviewed in detail in ref. [8], theoretical speculations about the existence of strangelets may be summarized as follows:

1. *It is unclear whether bulk strange quark matter exists at all.*

2. *It is unclear whether bulk strange quark matter can be stable.* If it does exist, strange quark matter may be absolutely stable in bulk at zero external pressure, though the expected values for the relevant parameters tend to disfavour stability [1].

3. *Finite size effects make it very unlikely that small strangelets (A < 10) can be stable or long-lived.* Even if bulk strange quark matter is stable, finite-size effects (surface tension and curvature) significantly destablize strangelets with low baryon number. For typical parameters, it has been estimated that finite-size effects add, e.g., 50 MeV per baryon for A = 20 and 85 MeV per baryon for A=10 [1].

4. *Stable strangelets, if they exist, could be present only in states of low entropy (i.e., temperature).* Hot strangelets are much less stable than cold ones. The characteristic scale to decide what is hot or cold is set by the binding energy per baryon of the strangelet. On general thermodynamic grounds, the time-scale for evaporation of hot strangelets is expected to be very small, though difficult to calculate. Assuming typical nuclear binding energies of O(1) MeV, one expects that stable strangelets need to be much colder than the matter produced in heavy-ion collisions [1,8].

5. *If stable strangelets exist, they are most likely positively charged.* If strange matter contained equal numbers of *u*, *d* and *s* quarks, it would be electrically neutral. Since, *s* quarks are heavier, Fermi gas kinematics alone indicates that strange quarks are suppressed, giving strange matter a positive charge per unit baryon number. However, the effects of gluon exchange reactions are difficult to quantify. Perturbatively, gluon exchange is repulsive and increases the mass. But gluon interactions weaken as quark masses are increased, so the gluonic repulsion is smaller between *s-s*, *s-u* or *s-d* pairs than between *u* and *d* quarks. Hence, increasing the strength of gluon interactions makes the charge of quark matter negative, but it also unbinds it. Unreasonably low values of the bag constant are necessary to compensate for a large repulsive gluonic interaction energy, which is why negatively-charged strangelets are regarded as extremely unlikely [8].

Hypothetical disaster scenarios based on strangelet production in heavy-ion collisions require that strangelets be stable or very long-lived, and hence that they are sufficiently cold. It has been argued in detail [8] that each of these conditions is unlikely. In the following, we discuss in particular how RHIC data allow us to strengthen the argument that sufficiently cold strangelets cannot be produced in the hot particle furnace created in a heavy ion collision. Moreover, most hypothetical disaster scenarios require that the produced strangelet be charged negatively, so that its fusion with positively charged nuclei could lead to a hypothetically disastrous chain of events. In contrast, in normal matter, positively-charged strangelets would capture electrons, which would shield any fusion with other nuclei. To trigger a run-away reaction in the latter case, one must invoke an ionizing mechanism, e.g., by transporting the strangelets to the interior of the Sun [12], and this adds another layer of unlikely assumptions.

**Strangelet production mechanisms in heavy-ion collisions**

Strangelets, if they exist at all, are hadronic systems made out of quarks. Any model for their production should be first tested against the existing data on the production of nuclei. This line of argument has been explored in ref. [1]. Here, we sharpen its conclusions in the light of recent data from RHIC. There are three models of particle production, which have been considered in the context of strangelet production in heavy-ion collisions.

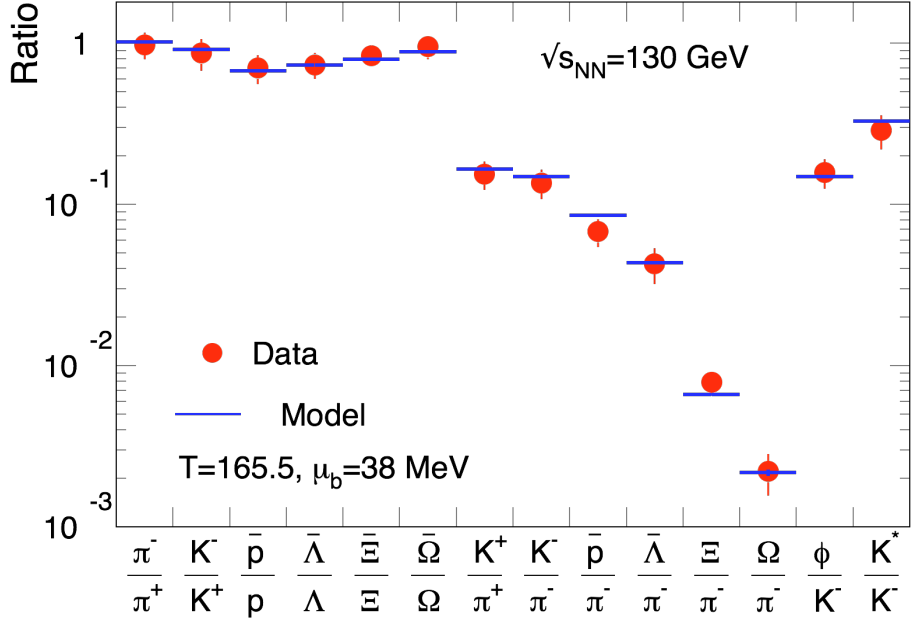

*Fig. 3: Comparison of the experimental data on different particle multiplicity ratios obtained at RHIC in Gold-Gold collisions at $\sqrt{s_{NN}}$ = 130 GeV with thermal model calculations [23]. The abundances of strange and multiple strange hadron species are well-described in terms of a chemical freeze-out temperature and baryon chemical potential $\mu_B$. We note that the freeze-out temperatures here and in Fig. 2, corresponding to $\sqrt{s_{NN}}$ = 200 GeV, are very similar, whereas the baryon chemical potential decreases with $\sqrt{s_{NN}}$. Other data on the dependences of these parameters on the center-of-mass energy are shown in Fig. 4.*

1. **Thermal models**

Hadron production in heavy-ion collisions is remarkably well described in terms of a statistical model. This model describes hadron yields in terms of the grand canonical ensemble of a hadron resonance gas at temperature $T$ and baryon chemical potential $\mu_B$ [3], which characterizes the net baryon density. Figures 2 and 3 illustrate the success of this model for particle production in heavy-ion collisions at RHIC. The relative abundance of all particle species with $n_s$ strange quarks (where experimental observations extend up to $n_s$ = 3 only), are well described by the model. The temperature $T$ and baryochemical potential $\mu_B$ of this model show a characteristic dependence on the center-of-mass energy of the heavy-ion collision, as seen in Fig. 4. The temperature increases with increasing collision energy, saturating at $T \approx 165$ MeV, whereas the baryon chemical potential $\mu_B$ decreases. This is reflected in the reduction as the center-of-mass energy increases in the rate of production of $\Omega$ baryons, which contain three strange quarks, as can be seen by comparing Figs. 2 and 3. The reason for the decrease of $\mu_B$ is that, at higher collision energies, the same net baryon number is distributed over a wider longitudinal kinematic range, resulting in a lower net baryon density and hence a lower value of $\mu_B$.

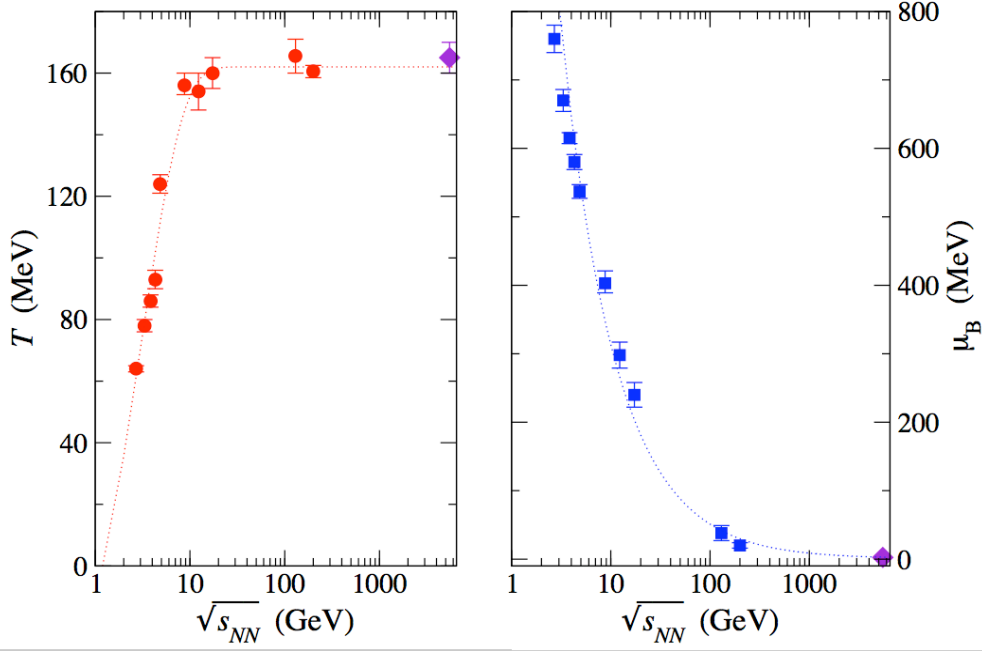

*Fig. 4: Thermal model fits at mid-rapidity of the freeze-out temperature T and the baryon chemical potential $\mu_B$ as functions of the center-of-mass energy $\sqrt{s_{NN}}$. The data points up to RHIC energies are taken from ref. [23]. The points at $\sqrt{s_{NN}}$ = 5.5 TeV are based extrapolations of the measured trend [24]. The decrease of baryon chemical potential with center-of-mass energy makes strangelet production less likely at higher center-of-mass energies.*

The statistical approach can also be applied to the production of complex nuclei. The penalty factor for the yield $Y_A$ of a nucleus $A$ compared to $A$-1 is the ratio of Boltzmann weights [1]

$$P_F = \frac{Y_A}{Y_{A-1}} \approx \exp\left[-\frac{(m_N - \mu_B)}{T}\right], \qquad (1)$$

where $m_N$ is the nucleon mass. For the relevant temperature of $T$ = 165 MeV, a small baryon chemical potential $\mu_B \ll m_N$ as shown in Fig. 4, and $A$ = 10, this gives a relative suppression factor $3 \times 10^{-25}$ compared to the production of nucleons. We note that this would be the suppression factor for the production of normal nuclei. The production of any strangelet of $A$ = 10 may be expected to be suppressed significantly more than the production of a normal $A$ = 10 nucleus. Moreover, the grand canonical ensemble, on which the above estimates are based, is expected to be modified by an additional canonical suppression factor, as soon as the constraint from the finite total baryon number in the collision becomes relevant at sufficiently large $A$. Taking these considerations into account, the suppression factor $3 \times 10^{-25}$ is an extremely conservative upper bound. If one repeats the exercise with $A$ = 20, one is led to a suppression factor $10^{-49}$.

We note that the production of light nuclei with $A \ll 10$ has been measured in central heavy-ion collisions at AGS, SPS and RHIC. It is well accounted for by the penalty factor in eq. (1): see, e.g. ref. [25] for a comparison of thermal model calculations to data. The measured penalty factors for light nuclei range from 1/50 at AGS, to about 1/300 at the SPS, and for antinuclei they range from $1/(2 \times 10^5)$ at the AGS and 1/3000 at the SPS to 1/1500 at RHIC [1]. As functions of center-of-mass energy, the penalty factors increase for nuclei but, due to the decrease of the baryon chemical potential with $\sqrt{s_{NN}}$, they decrease for antinuclei. In a system with postive net baryon number, the total yield of nuclei is always larger than the yield of antinuclei. For this reason, the above estimate is based on the thermal production of nuclei.

## 2. Coalescence models

The basic physics idea of coalescence models is that a nucleus $A$ forms when $A$ nucleons occupy the same interaction volume. In these models, the yield $Y_A$ of nuclei $A$ is related to the initial yield $Y_N$ of nucleons as

$$Y_A = B_A \left(Y_N\right)^A, \qquad (2)$$

where $B_A$ is the so-called coalescence parameter. The penalty factor $P_F$ for coalescing an additional nucleon onto an existing cluster is then [1]

$$P_F = \frac{Y_A}{Y_{A-1}} = \left(\frac{B_A}{B_{A-1}}\right)_0 \frac{V_0}{V} Y_N. \qquad (3)$$

Here, $V$ denotes the interaction volume over which coalescence is effective, and the subscript '0' refers to a reference scale set, e.g., by determining the coalescence parameter and the interaction volume at a specific collision energy.

It has been emphasized previously in ref. [1] that predictions from the coalescence models are in qualitative and even reasonable quantitative agreement with thermal models. For instance, in ref. [1] it was estimated on the basis of coalescence models that the suppression factor for production of an $A = 20$ nuclei in a central heavy ion collision is between $10^{-53}$ and $10^{-46}$. This compares very well with the suppression factor of order $10^{-49}$ obtained in the above discussion of thermal models.

Since coalescence models do not differ qualitatively from thermal models, the same safety arguments apply. For this reason, we emphasize in the text that in the detailed study of heavy-ion collisions at RHIC and lower energies, no evidence for an anomalous coalescence mechanism has been found. The basis of the 2003 report has been fully vindicated by further RHIC running.

## 3. Distillation mechanism

Strangeness distillation has been proposed specifically as a mechanism for strangelet production. This mechanism assumes that a baryon-rich quark-gluon plasma is produced in a heavy-ion collision, which cools by evaporation from its surface. Due to the large baryon chemical potential in this plasma, an $\bar{s}$ antiquark would be more likely to pair with a $u$ or $d$ quark, than an $s$ quark with a $\bar{u}$ or $\bar{d}$ antiquark. As a consequence, the cooling of the plasma would lead to an excess of $s$ quarks in a baryon-rich lump, which may finally become a strangelet.

We note that this production process would be more likely for large baryon chemical potential, and thus would be less likely for heavy-ion collisions at the LHC than at lower center-of-mass energies. Moreover, there is by now significant empirical evidence against a dynamical picture of heavy-ion collisions in which strangeness distillation could be operational. In particular, empirical evidence from RHIC strongly supports explosive production scenarios, in which, for instance, collective-flow gradients increase with center-of-mass energy [26]. The short lifetime of the produced systems (of the order of 10 fm/c) is not expected to allow for an evaporation process. Moreover, the explosive collective dynamics is expected to favor bulk emission rather than surface emission [26]. So, there is no evidence for a distillation mechanism capable of strangelet production at RHIC, and this suggestion for strange particle production has been abandoned for the LHC.

**Direct experimental searches for strangelets**

Strangelets have been searched for in ordinary matter on Earth [27] and in heavy-ion collisions over a wide range of center-of-mass energies. In particular, searches for stra
lets have been reported by several experiments at the Brookhaven Alternating Gradient Synchrotron [28], by the NA52 Collaboration at the SPS [29], and by the STAR Collaboration at RHIC [15]. All of these searches yielded negative results and reported complementary upper limits. In particular, STAR reported an upper limit of less than $10^{-6}$ strangelets per central Au-Au collision for strangelets with lifetimes larger than 0.1 ns and mass larger than 30 GeV. More details about the experimental situation can be found in the previous reports [1,8].

**Summary of the safety argument**

**1. Quantitative considerations**

The maximal luminosity of lead-lead (Pb+Pb) collisions at the LHC is $L = 10^{27}$ cm$^{-2}$ s$^{-1}$. With a hadronic Pb+Pb cross section of 8 barn, this implies a rate of up to 8000 Pb+Pb collisions per second. With a foreseen running time of 1 month per year ($10^6$ seconds) times a duration of the program of, say, 10 years, we arrive at a conservative upper bound on the total number of ion-ion collisions at the LHC of $O(10^{11})$. However, a large fraction of the hadronic Pb+Pb cross section is diffractive or very peripheral. Only 10 percent of the entire rate can be considered as being sufficiently central for creating a collision system characteristic of a heavy-ion collision with a number of participants $N_{part}$ > 20. As a consequence, a conservative bound on the number of heavy ion collisions relevant for production of an $A = 10$ nucleus is $O(10^{10})$.

Our conservative estimate for the thermal production of a *normal A = 10* nucleus at the LHC was $3 \times 10^{-25}$ times the rate of nucleon production. Taking the latter rate to lie in the hundreds, we arrive at a probability of $10^{-13}$ that a single normal nucleus of size $A = 10$ is produced during the entire LHC program as a result of the essentially thermal dynamics in a heavy ion collision. So, if LHC would run for the entire lifetime of the Universe, the probability of producing such a single nucleus via thermal production would be 1/1000[1].

We note that the above is an estimate for the thermal production of a *normal A = 10* nucleus from a hadron gas of temperature $T = 165$ MeV. The production of normal nuclear matter provides an extremely conservative upper bound on the production of strange quark matter. For this reason, we find that the significant empirical support for thermal particle production in heavy ion collisions, which was substantiated further by RHIC data in recent years, strengthens the main conclusion of the 2003 report [1]. There is no basis for any conceivable threat from strangelet production at the LHC.

---

[1] One may add that in semi-peripheral collisions, nuclei with $A = 10$ may appear amongst the break-up products of the spectators of the nuclear projectile. However, such fragment production of nuclear remnants is not a mechanism that could give rise to strangelets. For this reason, we focus solely on thermal production rates of normal nuclei.

## 2. Qualitative considerations

The above estimate of an upper limit to the probability of $A = 10$ nuclei can be further strengthened by the following qualitative argument, which is based on general principles of thermodynamics alone.

Strangelets are cold, dense systems. Like nuclei, they are bound by O(1-10) of MeV (if they are bound at all). Heavy-ion collisions produce hot systems. At LHC, the temperatures reached are in excess of 100 MeV. The second law of thermodynamics fights against the condensation of a system an order of magnitude colder than the surrounding medium. The hypothetical production of a cold strangelet from a hot hadron gas has been compared to producing an ice cube in a furnace [8].

This paper has aimed at communicating this central qualitative idea. In this appendix, we have provided the quantitative background, to which the notion of "particle furnace" corresponds. As seen from Fig. 4, measurements show that heavy-ion collisions reach temperatures of $T = 160$ MeV in the last stage of the collision. Moreover, the baryon chemical potential, which characterizes the net quark density, decreases as the center-of-mass energy increases, further decreasing the likelihood of producing any system with large atomic number. The particle abundances measured at RHIC and in lower-energy experiments are consistent with expectations from the thermal model of statistical hadronization (see Fig. 3). This model is also known to apply to the production of light nuclei, as far as they have been identified experimentally, and it provides a very large suppression factor for the production of $A = 10$ nuclei. On these grounds, we conclude that the experimental evidence from RHIC for the thermal model of particle production significantly strengthens the conclusions of the 2003 report of the LHC Safety Study Group.